\documentstyle[12pt]{article}
\def\lqq{\lq \lq }

\begin{document}
\begin{titlepage}

\rightline{ICN-UNAM 96-07}
\rightline{June 1996}
\rightline{hep-th/9606092}
\begin{center}
{\LARGE  Quasi-Exactly-Solvable Many-Body Problems}

\vskip 0.3cm
{\large by
\vskip 0.3cm
A. Minzoni}$^{\it a}$\\
{\it FENOMEC--IIMAS, UNAM, Apartado 20-726, 01000 MEXICO, D.F.}
\vskip 0.3cm
{\large M. Rosenbaum}$^{\it b}$,
{\large A. Turbiner}$^{\it c}$\\
{\it Instituto de Ciencias Nucleares, UNAM, Apartado Postal 70-543, 
04510 Mexico D.F., Mexico}
\vskip 0.3cm
\end{center}

\begin{center}
{\large ABSTRACT}
\end{center}
\vskip 0.3 cm
\begin{quote}
Explicit examples of quasi-exactly-solvable $N$-body problems 
on the line are presented. These are 
related to the hidden algebra $sl_N$, and they are 
of two types -- containing up to $N$ (infinitely-many eigenstates are
known, but not all) and up to 6 
body interactions only (a finite number of eigenstates is known).
Both types degenerate to the Calogero model.
\end{quote}

\vskip 2cm
\noindent
$^{\it a}$E-mail:  tim@uxmym1.iimas.unam.mx

\noindent
$^{\it b}$E-mail:  mrosen@roxanne.nuclecu.unam.mx 

\noindent 
$^{\it c}$On leave of absence from the Institute for Theoretical
and Experimental Physics,
Moscow 117259, Russia\\E-mail: turbiner@axcrnb.cern.ch,
turbiner@roxanne.nuclecu.unam.mx
\end{titlepage}

The Calogero model \cite{calo} is one of the most remarkable objects 
in non-relativistic multidimensional quantum mechanics. Moreover, 
a quite exciting relation of this model with to the two-dimensional 
Yang-Mills theory has been found recently \cite{gn}. 
The Calogero model has many 
beautiful properties such as: complete-integrability, 
maximal super-integrability and being an exactly-solvable 
$N$-body problem on the real line. 
The model is defined by the Hamiltonian
\begin{equation}
\label{e1}
H_{Cal} ={1 \over 2} \sum_{i=1}^N \left[ -\partial_i^2 + 
\omega^2 x_i^2 \right] + \sum_{j < i}^N \frac g {(x_i-x_j)^2} + V^*\ ,
\end{equation}
where $\partial_i \equiv \frac{\partial}{\partial x_i}$,  $V^*=0$, 
$\omega$ is the harmonic oscillator frequency, hereafter normalized 
to $\omega=1$, and
$g > -1/8$ is the coupling constant. Recently, it was 
found \cite{rt} that the Calogero model is  characterized by the 
hidden algebra $sl_N$ and that
the Calogero Hamiltonian (1) is a 
Lie-algebraic, exactly-solvable operator. The goal of this Letter is 
to show that the Calogero Hamiltonian can be generalized 
to a Lie-algebraic, {\it quasi-}exactly-solvable operator 
leading to explicit examples of quasi-exactly-solvable 
$N$-body problems.

For the above purpose, let us first make a gauge rotation 
in equation (1) taking the Calogero ground-state wave function 
as a gauge factor \cite{tur}. We have:
\[
h = -2\ \Psi_0^{-1}H_{Cal} \Psi_0 \equiv 
- 2 {\beta(x)^{-\nu}} e^{\frac{1}{2}\sum^{N} x_i^2} H_{Cal} 
{\beta(x)^{\nu}} e^{-\frac{1}{2} \sum^{N} x_i^2
} =
\]
\[
=\  \sum_{i=1}^N  \partial_i^2 \ -\ 2 \sum_{i=1}^N  x_i \partial_i\ +\
\nu \sum_{j\neq i}^N \frac 1 {x_i - x_j} [\partial_i - \partial_j]
\]
\begin{equation}
\label{e2}
- N - \nu N(N-1) + 2 V^*\ , \nonumber
\end{equation}
where $\nu$ is one of two solutions to the equation $g=\nu(\nu - 1)$,
and
$\beta(x)=\prod_{i>j}(x_i-x_j)$ is the Vandermonde determinant.
For the sake of simplicity, from hereon we shall omit the constant term
in (2),
since it only shifts the reference point of the  spectrum.
As the next step, we introduce the translation-invariant elementary 
symmetric polynomials \cite{rt} :
\begin{equation}
\label{e3}
\tau_n (x) = \sigma_n(y(x)) \ , \ n=2,3,\ldots , N \ ,
\end{equation}
where $\sigma_n(z)$ are the standard elementary symmetric polynomials 
(see, for example, \cite{mac}), 
\begin{equation}
\label{e4}
Y\ =\ \sum_{j=1}^N x_j\ ,\ 
y_i\ =\ x_i - \frac{1}{N} \sum_{j=1}^N x_j\ ,\quad i=1,2,\ldots , N ,
\end{equation}
with the constraint $\sum_{i=1}^N y_i=0$, and $Y$ is the center-of-mass 
coordinate. Making the change of variables\footnote{For this 
transformation the Jacobian in explicit form is not known so far}
\[
(x_1,x_2,\ldots x_N) \rightarrow (Y, \tau_n (x)|\ n=2,3,\ldots , N),
\]
the operator $h$, after extraction of the center of mass motion, 
transforms into (see [3]):
\[
h_{rel} = \sum^N_{j,k=2} A_{jk}
	\frac{\partial^2}{\partial\tau_j\partial\tau_k}
-2 \sum^N_{j=2} j \tau_j { \partial \over \partial\tau_j} - 
\left(\frac{1}{N} + \nu\right)
 \sum^N_{j=2} (N-j+2)(N-j+1)  \tau_{j-2} { \partial \over \partial\tau_j}
\]
\begin{equation}
\label{e5}
+V^*(\tau)\ ,
\end{equation}
where 
\[
	A_{jk} =  {(N-j+1) (k-1) \over N} \tau_{j-1} \tau_{k-1} +
	\sum_{\ell \geq {\rm max}(1,k-j)} (k-j-2\ell) \tau_{j+\ell-1}
	\tau_{k-\ell-1}
\]
Here we put $\tau_0=1, \tau_1=0$ and  $\tau_p=0$, if $p<0$ and $p>N$. 
It is worth noting that the Calogero Hamiltonian (1) is $Z_2$-invariant: 
$(x_i \rightarrow -x_i)$; \linebreak the $\tau$-variables are
(anti)symmetric under this transformation:\linebreak 
$\tau_n(-x_1,-x_2,\ldots ,-x_N) = (-)^n\tau_n(x_1,x_2,\ldots ,x_N)$.

Now let us introduce the real algebra $sl_N (\tau)$ of first-order 
differential operators in the most degenerate representation, where 
all spins vanish except one :
\[
J_i^- = { \partial \over \partial \tau_i}  \ ,   \quad i=2,3,\ldots , N \ ,
\]
\[
J_{i,j}^0 = \tau_i J_j^-=\tau_i { \partial \over \partial \tau_j} \ , \quad i,j=2,3,
\ldots , N \ ,
\]
\begin{eqnarray}
\label{e6}
J^0(n) = n - \sum_{p=2}^N \tau_p \frac{\partial}{\partial \tau_{p}} \ ,
\end{eqnarray}
\[
J_i^+(n) = \tau_i J^0\ , \quad i=2,3,\ldots , N \ .
\]
If the parameter $n$ is a non-negative integer number, the representation (6)
becomes finite-dimensional  and its representation space is given by
the space of polynomials
\begin{equation}
\label{e7}
{\cal P}_n \ =\ \mbox{span} \{ \tau _2^{n_2} \tau _3^{n_3} \tau_4^{n_4} 
\ldots \tau_{N}^{n_{N}} : 
0 \leq \sum n_i \leq n \}\ .
\end{equation}
It is worth noting that $J_{i,j}^0$ form the algebra $sl_{N-1} \subset sl_N$ 
and, in turn, $J_i^-, J_{i,i}^0$ form the sub-algebra $b_2 \subset sl_N$.

It is evident that when $V^*(\tau)=0$ the operator $h_{rel}$ can be 
rewritten in terms of the generators of $sl_N(\tau)$ given by (6) and, furthermore, it does not contain the generators $J_i^+$. This 
implies that the operator $h_{rel}$ is exactly-solvable, i.e. this
operator preserves 
the flag of spaces ${\cal P}_n: {\cal P}_0 \subset {\cal P}_1 \subset 
{\cal P}_2 \ldots$ 
(see \cite{ams} and also \cite{rt}). 

We now proceed to study the eigenfunctions of the operator $h_{rel}$
given in (5). Because 
$A_{2 2} = - 2 \tau_{2}$ depends on $\tau_{2}$ only, then, for the case
$V^{*} = V^{*} (\tau_{2})$, we have the remarkable property of the existence
of a family of eigenfunctions of $h_{rel}$ depending on $\tau_{2}$ only.
Due to this property the original eigenvalue problem 
\begin{equation}
\label{e8}
h_{rel} \phi (\tau_2) = \epsilon  \phi (\tau_2)\ ,
\end{equation}
is simplified to
\begin{equation}
\label{e9}
h_{rel}(\tau_2) \phi (\tau_2) = \epsilon \phi (\tau_2)\ ,
\end{equation}
where
\begin{equation}
\label{e10}
h_{rel}(\tau_2) = -2\tau_2 \frac{\partial^2}{\partial\tau_2^2}
-\left(4 \tau_2 + b_2\right) { \partial \over \partial\tau_2} 
+ 2 V^*(\tau_2)\ .
\end{equation}
Or, in terms of the generators (6),
\begin{equation}
\label{e11}
h_{rel}(\tau_2) = - 2J_{2,2}^0 J_{2}^- - 4 J_{2,2}^0 - b_2 
J_{2}^- + 2 V^*(\tau_2)
\end{equation}
with 
\[
b_2=(1+\nu N) (N-1) \ .
\]

For the case of the Calogero model, $V^*=0$, it is easy to find the 
eigenfunctions and eigenvalues of (9)--(10) in the form:
\begin{equation}
\label{e12}
\phi^{\{k\}} (\tau_2)=L_k^{(\frac{b_2}{2}-1)}(-2 \tau_2)\ ,
\ \epsilon_k=-4k\ ,\ k=0,1,2,\ldots ,
\end{equation}
where $L_k^{(\alpha)}$ are the associated Laguerre polynomials.

In order to carry out a Lie-algebraic analysis of $h_{rel}(\tau_2)$
in equation (10), note that 
the algebra $sl_N (\tau)$ contains the sub-algebra $b_2 (\tau_2)$
formed by $J^{0}_{2,2}, J^{-}_{2}$: $b_2 (\tau_{2}) \subset sl_N (\tau)$.
This sub-algebra $b_2 (\tau_2)$ can be extended to $sl_{2} (\tau_{2})$ 
with generators given by
\begin{equation}
\label{e13}
J^{+} (n) = \tau^{2}_{2} {\partial \over \partial \tau_{2}} - n \tau_{2}\ ,\ 
J^{0} (n) = \tau_{2} {\partial \over \partial \tau_{2}} - {n \over 2}\ ,\ 
J^{-} = {\partial \over \partial \tau_{2}}\ ,
\end{equation}
such that for $n=0:\ J^0 (0) = J^{0}_{2,2}, J^{-}_{2} = J^{-}$. 
It is worth emphasizing that in this realization 
$sl_2 (\tau_{2}) \not\subset sl_N (\tau)$.
So in terms of the generators (13), the operator (10) takes the form 
\begin{equation}
\label{e14}
h_{rel}(\tau_2) = - 2 J^0(n) J^- - 4 J^0(n) - 
\left(b_2 + n \right) J^- - 2n + 2 V^*(\tau_2)\ ;
\end{equation}
(cf.(11)).

One can now pose the following natural question: could we gauge-rotate 
$h_{rel}(\tau_2)$ in the $\tau_2$ direction and fit  $V^*(\tau_2)$ 
in such a way as to obtain a Lie algebraic operator. In concrete terms 
this means that we want to find $\Psi^{*}(\tau_2)$ and $V^{*}(\tau_2)$ 
such that
\begin{equation}
\label{e15}
(\Psi^*(\tau_2))^{-1}h_{rel}(\tau_2)\Psi^*(\tau_2) \in U_{sl_2(\tau_2)} ,
\end{equation}
where $U_{sl_2(\tau_2)}$ denotes the universal enveloping algebra $sl_2$
taken in the representation (13). We have been able to find three concrete examples which provide an affirmative answer to this 
question and lead to three types of {\it quasi-exactly-solvable, many-body} problems.

(I). Take as a gauge factor $\Psi^*(\tau_2)=\tau_2^{\alpha}$ and 
choose $V^*=\frac{\gamma}{\tau_2}$. Then it is easy to see that 
the gauge-rotated operator $h_{rel}$ remains Lie-algebraic:
\[
h_{rel}^{(1)}(\tau_2) = \tau_2^{-\alpha}h_{rel}(\tau_2) \tau_2^{\alpha}
\]
\begin{equation}
\label{e16}
= - 2 J^0(n) J^{-}_{(n)} - 4 J^0(n) - \left(b_2 + n + 
4\alpha\right) J^{-}_{(n)} \ ,
\end{equation}
provided that
\begin{equation}
\label{e17}
2\gamma= \alpha \left(b_2 + 2\alpha -2\right)\ . 
\end{equation}
The resulting  modified Calogero Hamiltonian, which in this case 
contains up to $N$ body interactions, is given by: 
\begin{equation}
\label{e18}
H_{Cal}^{(1)} ={1 \over 2} \sum_{i=1}^N \left[ -\partial_i^2 + 
 x_i^2 \right] + \sum_{j < i}^N \frac g {(x_i-x_j)^2} + 
\frac{\gamma}{\tau_2}\ ,
\end{equation}
with the eigenfunctions 
\begin{equation}
\label{e19}
\Psi={\beta(x)}^{\nu} e^{-\frac{Y^2}{2}}{\tau_2}^{\alpha}
\left\{\begin{array}{c}
L_k^{(\frac{b_2}{2}+2\alpha - 1)} (-2\tau_2)\\
\phi_{\{k\}} (\tau_2,\tau_3,\ldots)
\end{array}\ ,
\right.
\end{equation}
and the eigenvalues
\begin{equation}
\label{e20}
\epsilon_k= -4k - 4\alpha \ ,
\end{equation}
(cf.(12)), corresponding to the eigenfunctions defined by Laguerre polynomials.
As before $\beta(x)$ is the Vandermonde determinant.
Unlike what occurs in the original Calogero model $(V^*=0)$, the remaining 
eigenfunctions $\phi_{\{k\}} (\tau_2,\tau_3,\ldots)$ 
are not polynomials anymore. Note in passing that for the particular case
$N=2$, the Hamiltonian $H_{Cal}^{(1)}$ becomes the well-known Kratzer Hamiltonian 
(see, for example, \cite{f}, problem 69).

A simple analysis shows that the requirement of normalizability 
(square-integrability at the origin) of the $\tau_2$-family 
of eigenfunctions (19) leads to the constraint
\begin{equation}
\label{e21}
\alpha > -\frac{b_2}{4} \ .
\end{equation}

Thus,  this deformation of the Calogero model allows us to find 
infinitely-many eigenstates explicitly but not all of them.
This situation is reminiscent of that occuring in the case of the 
Hulten and Saxon-Woods potentials, where the $s$-states can be found 
explicitly but not all other states
(see, for example, Flugge \cite{f}, problems 64, 68). 

(II). Another case leading to a truly quasi-exactly-solvable modification 
of the Calogero model can be constructed  by generalizing  the
previous example. Take as a gauge factor
\[ 
\Psi^{**}(\tau_2)=\tau_2^{\alpha} \exp(-\frac{a}{2}\tau_2^2-b\tau_2)
\] 
and choose $V^{**}=\frac{\gamma}{\tau_2}+A\tau_2^3+B\tau_2^2+C\tau_2$
with appropriate coefficients. 
Then the gauge-rotated operator $h_{rel}(\tau_2)$ remains 
Lie-algebraic: 
\[
h_{rel}^{(2)}(\tau_2) = 
\tau_2^{-\alpha} \exp(\frac{a}{2}\tau_2^2+b\tau_2)
h_{rel}(\tau_2) 
\tau_2^{\alpha} \exp(-\frac{a}{2}\tau_2^2-b\tau_2) 
\]
\[
= -2\tau_2 \frac{\partial^2}{\partial\tau_2^2}
+\left[4a\tau_2^2 + 4(b-1)\tau_2 - b_2\right] { \partial \over \partial\tau_2} 
-4an\tau_2-2n(b-1)
\]
\begin{equation}
\label{e22}
= - 2 J^0(n) J^- + 4a J^+(n) + 4(b-1) J^0(n) - 
\left(b_2 + n + 4\alpha\right) J^- \ ,
\end{equation}
(cf.(16)).
The corresponding modified Calogero Hamiltonian is of the form
\begin{equation}
\label{e23}
H_{Cal}^{(2)} ={1 \over 2} \sum_{i=1}^N \left[ -\partial_i^2 + 
 x_i^2 \right] + \sum_{j < i}^N \frac g {(x_i-x_j)^2} + 
\frac{\gamma}{\tau_2}+A\tau_2^3+B\tau_2^2+C\tau_2 ,
\end{equation}
provided that 
\[
A\ =\ a^2\ ,
\]
\[
B\ =\ 2a(b-1),
\]
\[
C\ =\ \left[ (b-1)^2 -1 -a\left(2n+1+2\alpha+ \frac{b_2}{2}\right)\right],
\]
with $\gamma$ given by (17). 
The eigenfunctions of this new Hamiltonian are:
\begin{equation}
\label{e24}
\Psi={\beta(x)}^{\nu} e^{-\frac{Y^2}{2}}{\tau_2}^{\alpha}
\exp(-\frac{a}{2}\tau_2^2-b\tau_2)
\left\{\begin{array}{c}
p_n^{\{k\}} (\tau_2)\ ,\ k=0,1,2,\ldots n\\
\phi_{\{k\}} (\tau_2,\tau_3,\ldots)
\end{array}\ ,
\right. 
\end{equation}
where the  $p_n^{\{k\}}$ are polynomials of degree $n$, while generically  
the $\phi_{\{k\}}$ are not polynomials.
In order to ensure the normalizability of (24), the parameter $\alpha$ 
should obey the constraint (21). Note that for the 
two-body case, the Hamiltonian  $H_{Cal}^{(2)}$ corresponds to one of 
the well-known examples of one-dimensional quasi-exactly-solvable problems
\cite{t}. For the general $N$-body case if $\alpha=0, b=1$ the 
Hamiltonian (23) coincides with that obtained in \cite{us}.
Let us emphasize that when $\alpha=0$ and, correspondingly 
$\gamma=0$, the quasi-exactly-solvable Hamiltonian 
(23) contains two, three, and up to six-body interactions only, 
independently on the number of bodies $N$. This follows immediately 
from the identity:
\[
2N \tau_2 = - \sum_{i>j} (x_i-x_j)^2 \ .
\]

The problem of finding the polynomials $p^{\{k\}}_n (\tau_2)$ in 
(24) is reduced to solving an algebraic equation of degree $n$ whose 
roots are the corresponding eigenvalues $\epsilon_n^{\{k\}}$. 
We just present the explicit formulae for 
$n = 0, 1, 2.$ For the sake of simplicity, we take $b = 1$.

It is clear that for $n = 0$ the polynomial eigenfunction is a constant,
$p_0^{\{0\}}=const$ and $\epsilon_0^{\{0\}}=0$.
For $n = 1$ the  two polynomial eigenfunctions have the form
\[ 
p_1^{\pm} = \tau_2 - \frac{\epsilon_1^{\pm}}{4a}\ ,
\]
where $\epsilon_1^{\pm}=\pm 2 \sqrt{a b_2}$ are the corresponding eigenvalues.
Note that $p_1^{\pm}(\epsilon_1^{\pm})$ form a double-sheeted Riemann 
surface in the parameter space $a(b_2)$ in complete agreement with \cite{t}.
For the case $n = 2$ the eigenfunctions are quadratic polynomials 
in $\tau_2$ and are given by
\begin{eqnarray}
\label{e25}
\nonumber
p^{\{1\}}_2 &=& \tau^2_2 - \frac{2 + b_2}{4a} \quad , \\
p_2^{\{0,2\}} &=& \tau^2_2 + \epsilon_2^{\{0,2\}} \tau_2 + {b_2
\over 4 a}\ ,  
\end{eqnarray}
with the corresponding eigenvalues
\begin{eqnarray}
\label{e26}
\nonumber
\epsilon_2^{\{1\}} &=& 0 \ ,\\
\epsilon_2^{\{0,2\}} &=& \pm 4 \sqrt{a (1 + b_2)} \ .
\end{eqnarray}
Observe that the number of zeroes of $p^{\{k\}}_2$, which is equal to $k$, 
agrees with the Sturm theorem.

(III). In order to proceed to another example of quasi-exactly-solvable 
many-body problems let us mention that since $A_{23}=-3\tau_3$ 
the original Calogero model (8) has, besides the eigenfunctions 
(12), another outstanding family of eigenfunctions of the form
\begin{equation}
\label{e27}
\phi^{\{k\}} (\tau_2,\tau_3) = \tau_3 L_k^{(\frac{b_2}{2}+2)}(-2 \tau_2)\ ,
\end{equation}
with eigenvalues
\[
\epsilon_k=-4k-6\ ,\ k=0,1,2,\ldots
\]
This functional form suggests to take as a gauge factor 
\[
\Psi^{***}(\tau_2,\tau_3)=\tau_3 \tau_2^{\alpha}\exp(-\frac{a}{2}\tau_2^2-b\tau_2) \ ,
\]
and the same functional form $V^{**}$ for the potential as in 
the previous section. Now we make the gauge 
transformation of the original $h_{rel}$ in (5) with this factor. 
Remarkably, we find that the resulting operator (after a suitable 
choice of the parameters in $V^{**}$) still has eigenfunctions depending on $\tau_2$ only ! Moreover, it is easy to see that the $\tau_2$-depending 
operator $h_{rel}^{(3)}(\tau_2)$, defining these $\tau_2$-depending 
eigenfunctions, is given by 
\[
h_{rel}^{(3)}(\tau_2) = 
\]
\begin{equation}
\label{e28}
= -2\tau_2 \frac{\partial^2}{\partial\tau_2^2}
+\left[4a\tau_2^2 + 4(b-1)\tau_2 - (b_2+6)\right]
\frac{\partial}{\partial\tau_2} - 4an\tau_2-2n(b-1)-6\ .
\end{equation}
 The Lie-algebraic form of (28) is 
\begin{equation}
\label{e29}
h_{rel}^{(3)}(\tau_2) = - 2 J^0(n) J^- + 4a J^+(n) + 4(b-1) J^0(n) - 
\left[(b_2 + 6) + n + 4\alpha\right] J^- -6\ ,
\end{equation}
(cf.(16), (22)).
The corresponding modified Calogero Hamiltonian $H_{Cal}^{(3)}$ coincides
with (23) with the following slight modifications in the parameters: 
$b_2$ is replaced by $b_2+6$, and the reference point of the spectrum 
is shifted by (-6). The corresponding eigenfunctions are then given by
\begin{equation}
\label{e30}
\Psi={\beta(x)}^{\nu} e^{-\frac{Y^2}{2}}{\tau_2}^{\alpha} \tau_3
\exp(-\frac{a}{2}\tau_2^2-b\tau_2)
\left\{\begin{array}{c}
p_n^{\{k\}} (\tau_2)\ ,\ k=0,1,2,\ldots n\\
\phi_{\{k\}} (\tau_2,\tau_3,\ldots)
\end{array}\ ,
\right. 
\end{equation}
(cf.(24)), 
where the  $p_n^{\{k\}}$ are polynomials of degree $n$, while 
generically the $\phi_{\{k\}}$ are again not polynomials.
The normalizability of (30) dictates that the parameter $\alpha$ 
should obey the constraint (21) with $b_2 \rightarrow (b_2+6)$. 
Also the expressions for $p_n^{\{k\}}$ obtained in (25)--(26) 
remain valid replacing $b_2 \rightarrow (b_2+6)$.
It is worth noticing that when $N=2$, the whole family of 
eigenfunctions (30) vanishes.

In conclusion, we showed the three cases of quasi-exactly-solvable 
many-body problems on the line characterizing up to $N$-body interactions 
(I), up to 6-body interactions (II)--(III) and having no $N=2$ limit (III).
All of them are associated with the Calogero model. These examples 
are described by the Hamiltonian: 
\[
H_{Cal}^{QES} ={1 \over 2} \sum_{i=1}^N \left[ -\partial_i^2 + 
 x_i^2 \right] + \sum_{j < i}^N \frac g {(x_i-x_j)^2} + 
\frac{\gamma}{\tau_2}+A\tau_2^3+B\tau_2^2+C\tau_2 ,
\]
where the parameters are
\newpage
\begin{equation}
\label{e32}
A\ =\ a^2\ 
\end{equation}
\[
B\ =\ 2a(b-1),
\]
\[
C\ =\ \left[ (b-1)^2 -1 -a\left(2n+1+2\alpha+ \frac{b_2}{2}+3\mu\right)
\right]
\]
\[
2\gamma= \alpha \left(b_2 +6\mu + 2\alpha -2\right)\ .
\]
The parameter $\mu$ can be either 0 or 1 leading to solutions of the 
form (24) or (30), respectively. The parameter $a$ is non-negative, 
while $b$ can be any real number, and  the 
parameter $\alpha$ is restricted to  satisfy 
\begin{equation}
\label{e33}
\alpha > -\frac{b_2+6\mu}{4} \ .
\end{equation}
If $a,b=0$, we return to the Hamiltonian (18) with solutions (19).

\end{document}